\documentclass[12pt,psfig,epsfig]{article}
\newcommand{\beq}{\begin{equation}}
\newcommand{\eeq}{\end{equation}}
\newcommand{\bea}{\begin{eqnarray}}
\newcommand{\eea}{\end{eqnarray}}
\def\lsim{\raise0.3ex\hbox{$\;<$\kern-0.75em\raise-1.1ex\hbox{$\sim\;$}}}
\def\gsim{\raise0.3ex\hbox{$\;>$\kern-0.75em\raise-1.1ex\hbox{$\sim\;$}}}
\def\para{\vspace{0.3cm}\noindent}

\begin{document}
\begin{center}
{\large \bf Prospect of determining the Dirac/Majorana state of neutrino  by Multi-OWL experiment }

\medskip

{Nayantara Gupta \footnote{Email address: nayan@imsc.res.in}and H.S. Mani \footnote{
Email address: hsmani@imsc.res.in }}\\
{\it Institute of Mathematical Sciences,\\
 C.I.T Campus, Taramani, Chennai 600113\\
  INDIA.}
  \end{center}

\begin{abstract}
We consider the non-radiative two body decay of a neutrino to a daughter neutrino with degraded energy and a very light particle (Majoron). Ultrahigh energy neutrinos from an astrophysical source like a Gamma Ray Burst undergoing this decay process are found to produce different number of events in the detector depending on whether they are Majorana or Dirac particles. The next generation large scale experiments like Multi-OWL is expected to provide us an accurate determination of the flux of neutrinos from astrophysical sources and this may enable us to distinguish between the Dirac and Majorana nature of neutrino.
  
\end{abstract}

PACS numbers:13.35.Hb,14.60.Lm 
\vskip 2cm

\section{Introduction}
  The recent obsevations from Super-Kamiokande \cite{sup}, KamLAND \cite{kam} and Sudbury Neutrino Observatory (SNO) \cite{sno} have established mixing between different neutrino flavours. There are, however, several questions which remain unanswered. The conventional model leads to three flavour states arising out of three mass eigenstates with $ \Delta m_{12}^2 \approx 7 \times 10^{-5} eV^2 $ and $ \Delta m_{23}^2 \approx 3 \times 10^{-3} eV^2 $.
 The Majorana/Dirac nature of the neutrino and the stability of all the states are still open questions. The Majorana/Dirac nature of neutrino can be resolved 
 unambiguously if neutrinoless double beta decay is observed. This difficult experiment is being carried out at various laboratories. Other alternative methods to establish Dirac/Majorana nature of neutrino would be desirable and this work is an attempt in that direction.

\para

We have been motivated by the recent work of Beacom et al. \cite{beacom1,beacom2}, who have considered the possibility that some of the neutrinos could be unstable. Depending upon the decay pattern they have noted that high energy neutrinos coming from astrophysical sources can have different flavour ratios. Such a decay of a ``heavy neutrino" to a ``light neutrino" could occur say by an emission of a spinless particle. For example, if neutrino mass pattern is given by normal hierarchy and if only the lightest neutrino is stable then the flavour ratio of the 
astrophysical neutrinos reaching the Earth would be $\nu_{e}:\nu_{\mu}:\nu_{\tau}=6:1:1$. 
An experiment like IceCube \cite{ice}, which could distinguish different neutrino flavours, may provide observational evidence for the different decay scenarios. 
\para

One has to consider detecting very high energy $(>10^{15}eV)$ neutrinos to
ensure the background from atmospheric neutrinos are negligible. Astrophysical
sources which could emit high energy neutrinos are Active Galactic Nuclei (AGN)
 and Gamma Ray Bursts (GRBs). Neutrino fluxes from such sources have been studied considering different theoretical models, and are expected to produce large number of events in experiments being planned ${\it e.g.}$ Antarctic Impulsive Transient Antenna (ANITA) \cite{anita}, Orbiting Wide-angle Light-collector (OWL) \cite{owl}. In this work, we explore the possibility of distinguishing the Dirac/Majorana nature of the neutrinos from these observations. The basic idea is this : If a ``heavy neutrino" decays to a ``light neutrino" with a spin flip then the light neutrino will
 be sterile if the neutrino is a Dirac particle. However, if the daughter
neutrino is a Majorana particle, this would still be an active neutrino and would produce a signal in the detector. We consider the neutrinos from a single Gamma Ray Burst. The nature of Gamma Ray Bursts, the emitted photon spectrum in the keV-MeV energy range and the subsequent emission of afterglow have been subjects of intense study in the last few years.       
Based on a ``fireball" model of GRB, the spectrum of high energy neutrinos from all GRBs has been obtained in \cite{waxman}. The expected neutrino spectrum from a single GRB in the same model has been calculated in \cite{nayan1,guetta}.

\para

 We show that the number of active neutrino events would be different in the detector depending on whether neutrinos are Dirac or Majorana particles. In particular we have looked at upward going tau neutrinos having energies above $3\times10^{15}eV$ taking into account of their regeneration as they traverse the Earth and look for signals of Dirac vs Majorana nature.
OWL \cite{owl} experiment will use the Earth's atmosphere as a detector. Multi-OWL \cite{owl1} would cover the Earth's horizon with a field of view of $120^{\circ}$ as it would view from an orbit of about $1000$ Km. The area on the Earth's surface, which would be within the field of view of Multi-OWL, would be used as the effective target area for the upward going neutrinos. 
Further, this experiment is suited for detecting tau neutrinos through observation of ultraviolet (UV) fluorescent light emitted by secondary charged particles.
 For energies above $10^{15}eV$ the main contribution would be from $\nu_{\tau}$ for the upward vertical showers as $\nu_{e}$ and $\nu_{\mu}$ would be absorbed in matter. The prospects of neutrino detection above energy $10^{15}eV$ by OWL experiment have been discussed in \cite{owl1}.

\para

In order to obtain results which do not depend sensitively on the model dependent parameters of the GRB, we have attempted to consider ratios of events belonging to different energy ranges. We discuss about the experimental accuracy needed
 to understand whether neutrino is a Dirac or Majorana particle based on the
event rates calculated for Multi-OWL experiment. 
\section{The flux of neutrinos from a single Gamma Ray Burst}
We consider a typical Gamma Ray Burst (GRB) whose parameters are inferred from observations, as our source of UHE (ultrahigh energy) neutrinos and calculate the number of tau neutrino events from it in Multi-OWL.
There are discussions on the parameters of a GRB and the mechanism of neutrino production in GRBs in the fireball model in APPENDIX A.
 We have used the neutrino spectrum from a single GRB in the fireball model \cite{guetta,nayan}.
The ultrahigh energy neutrino spectrum in the source (GRB) rest frame is as given below.
\beq
 \frac{dN_{\nu}(E_{\nu})}{dE_{\nu}}\propto \frac{1}{E_{\nu}^2}\left \{ \begin{array}
{l@{\quad \quad}l}
(E_{\nu}/E_{\nu}^{b})^{\alpha1}&E_{\nu}<E_{\nu}^{b} \\ (E_{\nu}/E_{\nu}^{b})^{\alpha2}&E_{\nu}^{b}<E_{\nu}<E_{\nu}^{s} \\ (E_{\nu}/E_{\nu}^{b})^{\alpha2}(E_{\nu}/E_{\nu}^{s})^{-2}&E_{\nu}>E_{\nu}^{s}
\end{array} \right.   
\eeq
where, $``E_{\nu}^b"$ is the neutrino spectrum break energy, $``\alpha1"$ and $``\alpha2"$ are the neutrino spectral indices below and above the neutrino spectrum break energy ($E_{\nu}^b$) respectively. $``E_{\nu}^b"$ is due to the break energy in the low energy photon spectrum of a GRB (see APPENDIX A for discussion).  $``E_{\nu}^s"$ is the break energy in the neutrino spectrum due to the synchrotron energy loss by high energy pions.   

\para

The final neutrino spectrum on Earth can be derived in the following way. Assuming that the GRB is occuring at a redshift $``z"$, the energy  per unit time and unit area to be measured by a detector on Earth is \cite{kolb}
\beq
F_{{\nu}obs}=\frac{L_{\nu}}{4\pi d^2_L}
\eeq

where, $``d_L"$ is the luminosity distance of the source.
The luminosity in the source rest frame $``L_{\nu}"$ can be expressed in terms
of the number of neutrinos per unit energy and unit time in that frame.
\beq
L_{\nu}=\int_{E_{{\nu}min}}^{E_{{\nu}max}} E_{\nu}\frac{dN_{\nu}^{\prime}(E_{\nu},t_{\nu})}{dE_{\nu}dt_{\nu}}dE_{\nu}
\eeq
where, $\frac{dN_{\nu}(E_{\nu})}{dE_{\nu}}=\int_{0}^{t_{d}} \frac{dN^{\prime}_{\nu}(E_{\nu},t_{\nu})}{dE_{\nu}dt_{\nu}}dt_{\nu}$,
 $``t_d"$ is the duration of the burst in the source rest frame.
We denote the observed neutrino spectrum on Earth per unit area, unit energy and unit time by $\frac{dM^{\prime}(E_{{\nu}obs},t_{{\nu}obs})}{dE_{{\nu}obs}dt_{{\nu}obs}}$.
 From eq.(2) and (3) we can write
\beq
\int_{E_{{\nu}min,obs}}^{E_{{\nu}max,obs}}E_{{\nu}obs}\frac{dM_{{\nu}obs}^{\prime}(E_{{\nu}obs},t_{{\nu}obs})}{dE_{{\nu}obs}dt_{{\nu}obs}}dE_{{\nu}obs}
=\frac{1}{4\pi d^2_L}\int_{E_{{\nu}min}}^{E_{{\nu}max}} E_{\nu}\frac{dN_{\nu}^{\prime}(E_{\nu})}{dE_{\nu}dt_{\nu}}dE_{\nu}
\eeq
Hence using $E_{{\nu}obs}=E_{\nu}/(1+z)$ and $d_L=(1+z)d_z$,
\beq
 \frac{dM_{{\nu}obs}^{\prime}(E_{{\nu}obs},t_{{\nu}obs})}{dE_{{\nu}obs}dt_{{\nu}obs}}= \frac{1}{4\pi d^2_z}\frac{dN_{\nu}^{\prime}(E_{\nu},t_{\nu})}{dE_{\nu}dt_{\nu}}
\eeq
$``d_z"$ is the comoving distance.
The number of neutrinos received on Earth per unit area and unit energy is
\beq
\frac{dM_{{\nu}obs}(E_{{\nu}obs})}{dE_{{\nu}obs}}=\frac{1}{4\pi d^2_z}(1+z)\frac{dN_{\nu}(E_{\nu})}{dE_{\nu}}
\eeq
where, $dt_{{\nu}obs}=(1+z)dt_{\nu}$, $\frac{dM_{{\nu}obs}}{dE_{{\nu}obs}}=\int_0^{t_{d obs}}\frac{dM_{{\nu}obs}^{\prime}(E_{{\nu}obs},t_{{\nu}obs})}{dE_{{\nu}obs}dt_{{\nu}obs}} dt_{{\nu}obs}$ and $t_{d obs}=t_d(1+z)$.
 In a spatially flat universe we calculate the comoving distance of a source using $\Omega_{\Lambda}=0.73$, $\Omega_{m}=0.27$ and $H_o=71 km sec^{-1} Mpc^{-1}$
from \cite{wmap}.
The source spectrum contains $\nu_{e}$ and $\nu_{\mu}$ in the ratio $1:2$. 
As the electron and muon neutrinos propagate from the source to the observer, 
the electron neutrinos oscillate to muon neutrinos which in turn can oscillate
 to tau neutrinos. If we assume $U_{e3}=0$, the solar neutrino mixing angle ${\theta}_{\odot}$ to be $30^{\circ}$ and the atmospheric neutrino mixing angle ${\theta}_{atm}$ to be $45^{\circ}$, then due to the effect of neutrino oscillations alone, we expect an equal number of $\nu_{e}$, $\nu_{\mu}$ and $\nu_{\tau}$ on Earth. Inclusion of the neutrino decay scenario causes the flavor ratio of the 
three active neutrino species to deviate from 1:1:1 \cite{beacom1}. 
We assume the decay scenario of neutrinos where one neutrino decays to
a daughter neutrino and a very light or massless particle.
The daughter neutrinos do not disappear and the neutrino mass spectrum is
hierarchical. The energy dependent neutrino decay term is proportional to $exp(-dm_{\nu}/E_{\nu}t)$, where $``d"$ is the distance of the source from the detector, $``m_{\nu}"$ is mass of neutrino and $``t"$ is its rest-frame lifetime. We assume the neutrino decays to be complete when we receive them on Earth. This assumption is a reasonable one because there is a lower limit on the value of $d/E_{\nu}$ set by the shortest distances of the sources (typically hundreds of Mpc) and the maximum energy  that will be visible in OWL. The daughter neutrinos would be slightly deflected from the direction of propagation of the parent neutrinos. Since, we assume that the source is emitting isotropically, the flux recieved on Earth would not change due to the angular deflection of the decay products.
If neutrinos are Dirac particles then daughters, which flip helicity, are 
sterile and undetectable. In the case of Majorana neutrinos, neutrinos and antineutrinos of both the helicity states are active and there is no distinction between a neutrino and an antineutrino.
The expression for the neutrino spectrum after the appearance of daughter neutrinos with degraded energies is given in eqn.(7) of \cite{beacom1}.
The energy dependence of the decay rate of neutrinos has been explicitly
given in Appendix B. $W(E^{\prime},E)$ is the energy distribution function of
 the daughter neutrinos where, $E^{\prime}$ and $E$ are the energies of the parent and daughter neutrinos respectively. This distribution function is different for Dirac and Majorana nature of neutrinos because in the Dirac case the neutrinos with flipped helicities are sterile in contrast to the Majorana neutrinos.
  We have also verified that in the case of radiative decay of neutrinos the energy dependence of the decay rate remains identical to the scalar decay scenario of neutrinos if one considers a coupling of the type $\sigma_{\mu \nu} F^{\mu \nu}$.
\section{Propagation of UHE neutrinos through the Earth}
The propagation of $\nu_{\tau}$ through the Earth's matter at very high
 energy does not diminish their number \cite{hal}. The vertically upward going $\nu_{e}$s and $\nu_{\mu}$s are absorbed in Earth's matter above $10^{16}eV$ and can not produce signals in the detector.
This is because the lifetime of $\mu$ lepton is large $(10^{-6}sec)$ and it loses all its energy before decaying. This is in contrast to a $\tau$ lepton whose lifetime is $10^{-13}sec$, hence it decays much before energy loss occurs due to electromagnetic  interaction.
 The $\nu_{\tau}$s produce $\tau$ leptons in charged  current interactions with matter nuclei. A $\tau$ lepton can decay in many ways but in all of these decay processes another $\nu_{\tau}$ is produced in the final state. The $\nu_{\tau}$ regenerated in this way carries one third  of the energy of the primary $\nu_{\tau}$ \cite{beacomr}.
Also there are neutral current interactions and the $\nu_{\tau}$ generated
in these interactions carries about half of the primary $\nu_{\tau}$ energy \cite{hal}. High energy $\nu_{\tau}$s initiate a series of interactions in the Earth and ultimately there is generation of $\nu_{\tau}$s of lower energies. Once the  mean free path of interaction of $\nu_{\tau}$ with nucleon of Earth's matter becomes comparable to the radius of the Earth, it directly propagates to the detector. There are discussions on propagation of tau neutrinos through the Earth and their event rates in detectors in a series of recent papers \cite{naumov,sharada,bottai,tseng,bugaev,yoshida}.  
 The procedure we have followed to calculate the probability of $\tau$ 
lepton emission from the surface of the Earth has been discussed in APPENDIX C.
The area on the surface of the Earth, which would be covered by the field of view of Multi-OWL, would act as the effective target area for the upward going tau neutrinos. One may look up \cite{owl1} for detailed discussion on how the larger field of view of Multi-OWL compared to OWL experiment would increase the effective target area on the Earth's surface for the upward going neutrinos.
 The $\tau$ leptons emitted from the surface of the Earth would produce  UV fluorescent light in air while passing through the atmosphere. If they decay before reaching the atmosphere, there will be charged particles in their decay products which can produce UV fluorescent light in the atmosphere. UV fluorescent light will be emitted isotropically, hence OWL would be able to detect the
charged particles from any direction \cite{owl_n}.   
The GRB $\nu_{\tau}$ spectrum on Earth folded with the probability of $\tau$ emission from the surface of the Earth, calculated in APPENDIX C and the target area of the upward going neutrinos on the Earth's surface gives the number of $\tau$ signals in the detector.

\section{Results}
The GRB parameters that have been used in our calculations are the following: a Lorentz factor $\Gamma=300$, total energy emitted in neutrinos $10^{51}erg$, low energy photon luminosity $L_{\gamma}=10^{51}erg/sec$, neutrino spectral indices $\alpha1=1.5$ and $\alpha2=-0.3, 0.05, 0.3$ in eqn. (1), variability time $t_{v}=10^{-3}sec$, $\epsilon_{e}=0.5$, $\epsilon_{B}=0.05$.
The values of these parameters have been inferred from low energy $\gamma$-ray observations on GRBs.  We have calculated neutrino events in Multi-OWL from a GRB at a redshift $z=0.5$ and at a zenith angle $\theta=180^{\circ}$. 
Since, the neutrino spectrum break energy $E^{b}_{\nu}$ is $2.26\times10^{15}eV$ for the GRB parameters used in our calculation and we are considering the neutrino events in the detector above energy $3\times10^{15}eV$, the GRB neutrino spectrum above $E^b_{\nu}$ only contributes to the number of events in the detector.
 The numbers of neutrino events calculated for Multi-OWL above energy $3\times10^{15}eV$ are of the order of hundred. If we take the ratio of the numbers of events in the detector above different energies our results become independent of the parameters of the GRB ${\it e.g.}$  $\Gamma$, $z$, $\epsilon^b_{\gamma,MeV}$ and depend only on $\alpha_2$ \cite{plb}. We calculate the expected numbers of active neutrino events in the detector above energies $3\times10^{15}$ and $2.2\times10^{16}eV$ for the two cases: (i) neutrino is a Dirac particle, (ii) neutrino is a Majorana particle. The ratios of the numbers of events above different energies are useful quantities for our study. They can be compared in future with their measured values in experiments like Multi-OWL. All GRB parameters except $\alpha_2$ have been kept fixed for our results displayed in TABLE I, TABLE II, TABLE III and TABLE IV.
 
\vskip 1cm 
TABLE I: Ratio of the numbers of vertically upward going $\nu_{\tau}$ events in Multi-OWL above energies $3\times10^{15}eV$ and $2.2\times10^{16}eV$ in the detector with mass ratio of daughter to parent neutrino 0.001 and $\alpha_2=-0.3$.
\begin{center}
\begin{tabular}{|c|c|c|c|c|} 
\hline\hline
Unstable&Branchings&$Ratio$&$Ratio$& $(R^M_{\nu_{\tau}}-R^D_{\nu_{\tau}})/R^D_{\nu_{\tau}}$\\ 
        &     &     $R^D_{\nu_{\tau}}$&$R^M_{\nu_{\tau}}$&\\
\hline
\hline
$\nu_{2}$,$\nu_{3}$&irrevalent&27.53 &29.99 &0.09  \\
\hline
$\nu_{3}$&$B_{3\rightarrow 1}=1$&24.92 &25.38 &0.02 \\
\hline
$\nu_{3}$&$B_{3\rightarrow 2}=1$&25.84 & 27.04&0.05 \\
\hline
$\nu_{3}$&$B_{3\rightarrow 1}=0.5$&25.4 &26.26 &0.03\\
         &  $B_{3\rightarrow 2}=0.5$& & & \\ 
\hline
\hline
\end{tabular}
\end{center}
\vskip 1cm
In TABLE I the unstable neutrino eigenstates/eigenstate and the decay modes considered have been mentioned in first and second columns respectively. 
The third and fourth columns show the ratios of the numbers of neutrino events in the detector above energies $3\times10^{15}$ and $2.2\times10^{16}eV$ for Dirac and Majorana nature of neutrinos respectively. The fifth column displays by what fraction the ratios differ in the two cases. The ratios of the numbers of active neutrino
events in the detector for the Dirac and Majorana nature of neutrinos differ
by $9\%$ for the decay scenario when both $\nu_2$ and $\nu_3$ are
 unstable and by $2\%$ for $B_{3\rightarrow 1}=1$. We have varied $\alpha_2$ to study its effect on the ratios of the numbers of events above different energies and our results have been presented in the next two tables.
\vskip 1cm
TABLE II: Ratio of the numbers of vertically upward going $\nu_{\tau}$ events in Multi-OWL above energies $3\times10^{15}eV$ and $2.2\times10^{16}eV$ in the detector with mass ratio of daughter to parent neutrino 0.001 and $\alpha_2=0.05$.
\begin{center}
\begin{tabular}{|c|c|c|c|c|} 
\hline\hline
Unstable&Branchings&$Ratio$&$Ratio$& $(R^M_{\nu_{\tau}}-R^D_{\nu_{\tau}})/R^D_{\nu_{\tau}}$\\ 
        &     &     $R^D_{\nu_{\tau}}$&$R^M_{\nu_{\tau}}$&\\
\hline
\hline
$\nu_{2}$,$\nu_{3}$&irrevalent&27.41 &29.85 &0.09  \\
\hline
$\nu_{3}$&$B_{3\rightarrow 1}=1$&24.83 &25.28 &0.02 \\
\hline
$\nu_{3}$&$B_{3\rightarrow 2}=1$&25.73 & 26.93&0.05 \\
\hline
$\nu_{3}$&$B_{3\rightarrow 1}=0.5$&25.3 &26.15 &0.03\\
         &  $B_{3\rightarrow 2}=0.5$& & & \\ 
\hline
\hline
\end{tabular}
\end{center}
\vskip 1cm
TABLE II shows the ratios of the numbers of events in the detector above energies $3\times10^{15}$ and $2.2\times10^{16}eV$ for $\alpha_2=0.05$. 
\vskip 1cm 
TABLE III: Ratio of the numbers of vertically upward going $\nu_{\tau}$ events in Multi-OWL above energies $3\times10^{15}eV$ and $2.2\times10^{16}eV$ in the detector with mass ratio of daughter to parent neutrino 0.001 and $\alpha_2=0.3$.
\begin{center}
\begin{tabular}{|c|c|c|c|c|} 
\hline\hline
Unstable&Branchings&$Ratio$&$Ratio$& $(R^M_{\nu_{\tau}}-R^D_{\nu_{\tau}})/R^D_{\nu_{\tau}}$\\ 
        &     &     $R^D_{\nu_{\tau}}$&$R^M_{\nu_{\tau}}$&\\
\hline
\hline
$\nu_{2}$,$\nu_{3}$&irrevalent&27.31 &29.71 &0.09  \\
\hline
$\nu_{3}$&$B_{3\rightarrow 1}=1$&24.74 &25.19 &0.02 \\
\hline
$\nu_{3}$&$B_{3\rightarrow 2}=1$&25.63 & 26.82&0.05 \\
\hline
$\nu_{3}$&$B_{3\rightarrow 1}=0.5$&25.2 &26.05 &0.03\\
         &  $B_{3\rightarrow 2}=0.5$& & & \\ 
\hline
\hline
\end{tabular}
\end{center}

If we compare TABLE I, TABLE II and TABLE III, we find that the percentages by which the ratios of the numbers of events in the Dirac and Majorana case differ for different decay modes are independent of the value of $\alpha_2$
 in the range $-0.3 \le \alpha_2 \le0.3$. Also, the ratios of the numbers of neutrino events do not vary significantly as we vary $\alpha_2$.
We can possibly observe neutrino events from a large number of GRBs whose values of spectral index $\alpha_2$ are in a range such that $(R^M_{\nu_{\tau}}-R^D_{\nu_{\tau}})/R^D_{\nu_{\tau}}$ are equal for all of them. 
 In principle, if  $(R^M_{\nu_{\tau}}-R^D_{\nu_{\tau}})$ is greater than the sum of  the theoretical and experimental errors in the ratios ($R^D_{\nu_{\tau}}$ or $R^M_{\nu_{\tau}}$) and  we find that for most of the GRBs the experimentally measured ratios of the numbers of neutrino events above energies $3\times10^{15}$ and $2.2\times10^{16}eV$ are closer to $R^D_{\nu_{\tau}}$ or $R^M_{\nu_{\tau}}$, then we would be able to say whether neutrino is a Dirac or a Majorana particle.
 We have discussed about the percentages of error in our theoretical calculation and neutrino experiments in section 5.  

\para

We also study the effect of the mass ratio of daughter to parent neutrino on the number of events in the detector. We denote the masses of the daughter and parent neutrinos by $m_l$ and $m_h$ respectively. The area covered by Multi-OWL on the Earth's surface would be $3.3\times10^{13}m^2$ \cite{owl1}. Our results are displayed in TABLE IV.

\vskip 1cm
TABLE IV: Number of vertically upward going $\nu_{\tau}$ events in Multi-OWL from neutrinos having energies above $3\times10^{15}eV$, $\nu_{3}$ unstable, $B_{3\rightarrow 1}=0.5,B_{3\rightarrow 2}=0.5$ and $\alpha_2=-0.3,0.3$.
\begin{center}
\begin{tabular}{|c|c|c|c|c|} 
\hline\hline
$m_l/m_h$&Dirac&Majorana&Dirac&Majorana  \\ 
 & $\alpha_2=-0.3$ & $\alpha_2=-0.3$ &$\alpha_2=0.3$&$\alpha_2=0.3$\\
\hline
\hline
0.01&698&735&891&938\\
\hline
0.1&699&736&892&940\\
\hline
0.25&702&744&896&949\\
\hline
0.5&711&773&908&987\\
\hline
0.75&731&829&933&1058\\
\hline
\hline
\end{tabular}
\end{center}
The number of events increases with increasing neutrino mass ratios and this increase is more pronounced in case of Majorana neutrinos. 
\section{Error Analysis}
The values of the parameters of a GRB are determined from $\gamma$-ray and afterglow  observations on that GRB. These values inferred from observations certainly carry some errors due to the errors in the measured data by the experiments. They would be used in the theoretical prediction of the number of neutrino events from that GRB and the errors in the measurement of the values of the GRB parameters would propagate to the theoretical calculation of expected number of neutrino events. We have calculated the ratios of the numbers of neutrino events above different energies in the detector, which depend only on the value of the GRB parameter $\alpha_2$. If it is possible to determine $\alpha_2$ accurately from the low energy $\gamma$-ray spectrum emitted by the GRB then the error in our calculation would become negligible. For example, if the error in the measurement of $\alpha_2$, $\triangle \alpha_2$ is 0.005 or 0.001 then the theoretical values of the ratios of the numbers of neutrino events ($R^D_{\nu_{\tau}}$ or $R^M_{\nu_{\tau}}$) would contain $1\%$ or $0.2\%$ error respectively, 
as $\frac{\triangle R^D_{{\nu}_{\tau}}}{R^D_{\nu_{\tau}}}$ or $\frac{\triangle R^M_{{\nu}_{\tau}}}{R^M_{\nu_{\tau}}} = (\frac{3\times10^{15}}{2.2\times10^{16}})^{\triangle \alpha_2}-1$ .  
Experimentally obtained ratios of the numbers of the neutrino events above different energies from a large number of GRBs can be compared with their
 theoretically possible values $R^D_{\nu_{\tau}}$ or $R^M_{\nu_{\tau}}$.
Detection of neutrinos from a large number of GRBs would reduce the statistical error in the measurement of the ratios of the numbers of neutrino events from them. If the systematic error of the detector in the measurement of number of neutrino events is $4\%$ then the Dirac/Majorana nature of neutrino would be distinguishable in that detector for the decay scenario with both $\nu_2$ and $\nu_3$ unstable using GRBs with $\triangle \alpha_2\approx 0.001$ as sources of ultrahigh neutrinos.

\section{Conclusion}
 We have calculated the ratios of the expected numbers of vertically upward going $\nu_{\tau}$ events from a single GRB in Multi-OWL experiment above energies $3\times10^{15}$ and $2.2\times10^{16}eV$, including the oscillation and non-radiative two body decay of neutrinos. The ratios of the numbers of active neutrino events in the detector would be different depending on whether neutrino is a Dirac or a Majorana particle. These theoretically calculated ratios carry errors only from the error in the determination of $\alpha_2$, which is obtained from the low energy photon data of the GRB.
Measurement of the ratios of the numbers of neutrino events from a large number of GRBs would reduce the experimental statistical error. If the systematic error  in detecting neutrino events is $4\%$, and the error in determination of $\alpha_2$ is of the order of 0.001, then we expect the Dirac/Majorana nature of neutrinos to be revealed in Multi-OWL for the scenario of neutrino decay when both $\nu_2$ and $\nu_3$ are unstable.
\section{Acknowledgement}
HSM gratefully acknowledges the support from DAE-BRNS scheme at the Institute of Mathematical Sciences, Chennai.  The authors would also like to thank C. V. Narasimhan, V. Datar and C. V. K. Baba for helpful discussions. 
\newpage
\begin{center}
{\bf{APPENDIX A}}\\
 The parameters of a Gamma Ray Burst
\end{center}
GRBs are short lived bursts of $\gamma$-rays of durations milliseconds to several minutes. After
 a burst emission of X-rays, optical and radio waves continue for several
 hours to months. This is called the afterglow phase of a GRB.
The observed radiation is produced both during the burst phase and the afterglow phase.
In the fireball model GRBs are produced by the dissipation of the kinetic energy of a relativistic expanding fireball. 
 keV-MeV energy photons are expected to be produced by synchrotron shock accelerated electrons.   
A GRB occurs at a redshift $``z"$, it has a Lorentz factor $``\Gamma"$ and it is expected to emit energy of the order of $10^{51} erg$ in keV to MeV energy $\gamma$-ray emission.
The observed photon spectrum on the Earth from a GRB can be expressed as a broken power law. The data obtained by BATSE (Burst and Transient Source Experiment)
 can be fitted using the parametrisation given by

\beq
\frac{d n_{\gamma}}{d \epsilon_{\gamma,ob}} \propto \left\{ \begin{array}{r@{\quad \quad}l}
{\epsilon_{\gamma,ob}}^{-\alpha_2-1} & \epsilon_{\gamma,ob}<\epsilon_{\gamma,ob}^{b}\\ {\epsilon_{\gamma,ob}}^{-\alpha_1-1} & \epsilon_{\gamma,ob}>\epsilon_{\gamma,ob}^{b}
\end{array} \right.  
\eeq
$\epsilon_{\gamma,ob}$ is the photon energy and $\epsilon_{\gamma,ob}^{b}$ is the break energy of the photon spectrum as measured on Earth.
The break energy is the characteristic observed energy of synchrotron photons
 produced by shock accelerated electrons.
 We denote the luminosity of a GRB in keV to MeV energy photon emission by
$``L_{\gamma}"$. 
There is a detailed discussion on photon and neutrino production from GRBs in the fireball model in \cite{wax1}.
 As the fireball expands relativistic shells come out from it with Lorentz factor $\Gamma \sim 100-1000$. 
The variability time $``t_v"$ of a GRB is the time interval between two successive emission of shells from it. When two shells collide with each other their kinetic energy is transformed into internal energy of the fireball. The fireball expands over a range of radii and reverse shocks are generated by the interaction of the expanding shells with the surrounding medium. 
Protons can be accelerated to very high energies by Fermi mechanism in internal and reverse shocks in the fireball.
The low energy (keV-MeV) photon spectral break energy in the source rest frame 
 is $``\epsilon^{b}_{\gamma,MeV}"$, where $\epsilon^{b}_{\gamma,MeV}=\epsilon_{\gamma,ob,MeV}^{b}(1+z)$, and it depends on the equipartition parameters $\epsilon_e$ and $\epsilon_B$, low energy photon luminosity of the GRB $L_{\gamma}$, Lorentz factor $\Gamma$ and variability time $t_v$. $``\epsilon_e"$ is the fraction of proton's shock thermal energy transformed to electrons and $``\epsilon_B"$ is the fraction of proton's shock thermal energy carried by the magnetic field in the fireball model.
 It is assumed that the wind luminosity carried by internal plasma energy $L_{int}$, is related to the observed $\gamma$-ray luminosity through 
$L_{int}=L_{\gamma}/{\epsilon_{e}}$. This assumption is justified because the electron synchrotron cooling time is short compared to the wind expansion time  hence, electrons loose all their energy radiatively. The expression for photon spectral break energy can also be expressed as \cite{wax1} 
\beq
\epsilon_{{\gamma},MeV}^{b}\approx {\epsilon_{B}}^{1/2}{\epsilon_{e}}^{3/2}\frac{L_{{\gamma},52}^{1/2}}{{\Gamma}_{2.5}^{2}t_{v,-2}}
\eeq
where, $L_{{\gamma},52}=L_{\gamma}/10^{52}$, $\Gamma_{2.5}=\Gamma/10^{2.5}$ and $t_{v,-2}=t_v/10^{-2}$.
Inside the fireball protons interact with photons to produce pions which subsequently decay to produce neutrinos.
There is a threshold energy for pion production in proton-photon interaction. 
For the energy of the proton-photon interaction to exceed the threshold energy 
for the $\Delta$ resonance, the proton energy in the source rest frame must
satisfy the condition \cite{guetta},
 
\beq
E_{p}^{b}\ge (1.4/{\epsilon^{b}_{\gamma, MeV}}){\Gamma}^{2}_{2.5}\times 10^{7} GeV
\eeq
The reference frame which has been defined as ``observer's rest frame" in \cite{nayan}, is referred as ``source rest frame" in this paper.
 From the expression of proton spectrum break energy one can obtain the neutrino spectrum break energy. We assume that on the average $20\%$ of the initial proton energy goes to the photo-pion. We also assume that the four final state leptons produced in the decay chain $\pi^{+}\rightarrow \nu_{\mu} {\mu}^{+}\rightarrow \nu_{\mu} e^{+} \nu_{e} \bar \nu_{\mu}$ equally share the pion energy.
Then the  expression for neutrino spectrum break energy in the source rest frame is as follows 

 \beq
E_{\nu}^{b} =7\times 10^{5}\frac{\Gamma_{2.5}^{2}}{\epsilon_{\gamma,MeV}^{b}}
  GeV.
\eeq 

The high energy pions produced due to proton-photon interactions inside the
fireball are expected to radiate energy due to their synchrotron emission.
One can compare the synchrotron energy loss time of a high energy pion with its lifetime for decay. Radiative energy loss of pions becomes important when their
 decay lifetime is larger than their synchrotron energy loss time. 
A break energy in the neutrino spectrum $E_{\nu}^s$ would appear due to the
synchrotron radiation emission of the pions before they decay to produce neutrinos. Assuming that a neutrino or an antineutrino takes only one fourth of the pion energy one can write down the expression for $E_{\nu}^{s}$ from \cite{guetta}.
\beq
E_{\nu}^{s}=10^{8}\epsilon_{e}^{1/2}\epsilon_{B}^{-1/2}L_{\gamma,52}^{-1/2}\Gamma^{4}_{2.5}t_{v,-2} GeV
\eeq
The total energy emitted by the GRB in electron, muon neutrinos and antineutrinos is $E_{\nu,HE}$. 
\beq
E_{\nu,HE}=\int_{E_{{\nu}min}}^{E_{{\nu}max}} E_{\nu} \frac{dN_{\nu}}{dE_{\nu}} dE_{\nu}
\eeq
where, $``E_{{\nu}min}"$ and $``E_{{\nu}max}"$ are the minimum and maximum neutrino energy in the source rest frame.
In the fireball model protons can be accelerated to a maximum energy of $10^{21}eV$. Since, a neutrino on the average receives $5\%$ of the proton energy, the maximum neutrino energy in the source rest frame $``E_{{\nu}max}"$ is approximately $10^{20}eV$.

\newpage
\begin{center}
{\bf{APPENDIX B}}
   
   Derivation of the factor $W(E^{\prime},E)$

\end{center}

We assume the decay of a heavy neutrino $\nu_{h}$ of mass $m_h$ to a light neutrino $\nu_{l}$ of mass $m_l$ occurs through $ \nu_{h} \rightarrow \nu_{l} + J $, where $J$ is a massless scalar(Majoron). The decay is given by the interaction Lagrangian density
\beq
L_I = g_1 \overline{\nu_{l}}\nu_{h} J + g_2\overline\nu_l \gamma_5\nu_h J+ h.c.
\eeq
The energy of the parent neutrino is $E^{\prime}$ and daughter neutrino is $E$.
The total decay rate in the rest frame of the parent neutrino is
\beq
\Gamma_{total}^{0}=\frac{m_{h}^2-m_{l}^2}{16\pi m_{h}^3}[|Re g_1|^2(m_l+m_h)^2+|Im g_2|^2 (m_h-m_l)^2]
\eeq
 The above expression can be derived by standard methods \cite{kim}. In the 
laboratory frame the decay rate is $\Gamma_{total}(E^{\prime})=\frac{m_h}{E^{\prime}}\Gamma_{total}^{0}$.  
For Majorana neutrinos, we have contribution from both conserved helicity and flipped helicity amplitudes $\Gamma_{total}(E^{\prime})=\Gamma_{f}(E^{\prime})+\Gamma_{nf}(E^{\prime})$. 
 The differential decay rate is $\frac{d\Gamma_{total}(E^{\prime},E)}{dE}$.
The total decay rate can also be written as  
\beq
\Gamma_{total}(E^{\prime}) = {\int_{E_{min}}^{E_{max}}\frac{d\Gamma_{total}(E^{\prime},E)}{dE}dE}. 
\eeq
 where, $E_{min} = \frac{m_{l}^2}{m_{h}^2} E'$ and $E_{max}=E^{\prime}$. We work in the laboratory frame where $E,E^{\prime} >> m_{h},m_{l}$.
 We obtain the energy distribution function of daughter neutrinos in the case of Majorana neutrinos as follows.
\beq
  W_{Majorana}(E^{\prime},E) =\frac{1}{\Gamma_{total}(E^{\prime})}\frac{d\Gamma_{total}(E^{\prime},E)}{dE}.
\eeq
The above expression can be expressed as a function of neutrino masses.
\beq
W_{Majorana}(E^{\prime},E)= \frac{m_{h}^{2}}{(m_{h}^{2}-m_{l}^{2})E^{\prime}}, with \frac{m_{l}^{2}}{m_{h}^{2}}E^{\prime}\leq E \leq E^{\prime}.
\eeq

For Dirac neutrinos only the daughter neutrinos, which conserve helicity, are contributing to the active species. So we define

\beq
W_{Dirac}(E^{\prime},E)=\frac{1}{\Gamma_{total}(E^{\prime})}\frac{d\Gamma_{nf}(E^{\prime},E)}{dE}.
\eeq

and we obtain the energy distribution function as given below.
\beq
W_{Dirac}(E^{\prime},E)= {m_h}^2\frac{m_{h}^{2}E-m_{l}^{2}E^{\prime}}{(m_{h}^{2}-m_{l}^{2})^2{E^{{\prime}2}}}
, with\frac{m_{l}^2}{m_{h}^2}E^{\prime} \leq E \leq E^{\prime}.
\eeq

Radiative decay going through transition magnetic moment will also lead to a similar result \cite{amit}.
\newpage
\begin{center}
{\bf{APPENDIX C}}
 
 Derivation of the probability $Prob({\nu_{\tau}}\rightarrow {\tau})$
\end{center}
The energy of the $\nu_{\tau}$ is denoted by $E_{\nu_{\tau}}(0)$ before it enters the Earth. While $\nu_{\tau}$ is propagating through the Earth it is losing energy by neutral current interactions. The expression for energy loss of $\nu_{\tau}$ by neutral current interaction is
\bea
\frac{dE_{\nu_{\tau}}(x)}{dx}=-\rho(x)N_A\int_{0}^{E_{\nu_{\tau}}(x)}
(E_{\nu_{\tau}}(x)-E_f(x))\frac{d{\sigma}_{NC}(E_{\nu_{\tau}}(x),E_f(x))}{dE_f(x)}dE_f(x)\nonumber\\
=-\rho(x)N_A{\sigma}_{NC}(E_{\nu_{\tau}}(x))(<E_f(x)>-E_{\nu_{\tau}}(x))\nonumber\\
=-\frac{E_{\nu_{\tau}}(x)}{2\lambda_{NC}(E_{\nu_{\tau}}(x))}
\eea
We have used the fact that in each neutral current interaction the $\nu_{\tau}$ loses half of its energy \cite{hal} in writing the final expression in eqn.(19).
$\sigma_{NC}(E_{\nu_{\tau}}(x))$ is the cross section and $\lambda_{NC}(E_{\nu_{\tau}}(x))$ is the mean free path of neutral current interaction. $E_f(x)$ is the energy of the neutrino after it loses energy by neutral current interaction.
The expression for the probability of neutral current interaction of $\nu_{\tau}$ is
\beq
P_{\nu_{\tau}\rightarrow \nu_{\tau}}^{0}(E_{\nu_{\tau}}(0),E_{\nu_{\tau}}(x),x) dE_{\nu_{\tau}}(x)=\delta(E_{\nu_{\tau}}(x)-E_{\nu_{\tau}}(0)e^{-\int_0^x\frac{dx^{\prime}}{2\lambda_{NC}(E_{\nu_{\tau}}(x^{\prime}))}})dE_{\nu_{\tau}}(x)
\eeq

The probability of a $\tau$ production with energy between $E_{\tau}(x)$ and $E_{\tau}(x)+dE_{\tau}(x)$ from a $\nu_{\tau}$ of energy $E_{\nu_{\tau}}(x)$ at a distance between $x$ and $x+dx$ is

\bea
P_{\nu_{\tau}\rightarrow {\tau}}^{0}(E_{\nu_{\tau}}(0),E_{\tau}(x),x) dE_{\tau}(x) dx&=& \rho(x) N_A dxdE_{\tau}(x)\int_{E_{{\nu}min}}^{E_{{\nu}max}}\frac{d\sigma_{CC}(E_{\nu_{\tau}}(x),E_{\tau}(x))}
{dE_{\tau}(x)} \nonumber\\
&\times&P_{\nu_{\tau}\rightarrow \nu_{\tau}}^{0}(E_{\nu_{\tau}}(0),E_{\nu_{\tau}}(x),x) dE_{\nu_{\tau}}(x).
\eea
where, $\sigma_{CC}(E_{\nu_{\tau}}(x))$ is the cross-section of charged current interaction for $\nu_{\tau}$ with energy $E_{\nu_{\tau}}(x)$, $\rho(x)$ is the density of the medium at $x$ and $N_A$ is Avogadro's number.
We assume that the $\tau$ lepton produced by charged current interaction carries $75\%$ of the $\nu_{\tau}$ energy \cite{beacomr}.
The $\tau$ leptons are losing energy during their propagation. There is discussion on energy loss of $\tau$ leptons in \cite{tseng}. The energy loss by $\tau$ leptons due to ionization is negligible compared to the other processes {\it e.g.} bremsstrahlung, $e^{+}e^{-}$ pair production and photonuclear 
processes for energies above $10^{5} GeV$. We have used the energy loss co-efficient of $\tau$ lepton
\beq
\beta(E_{\tau}(x))=(1.6+6(E_{\tau}(x)/10^{9}GeV)^{0.2})\times{10^{-7}}g^{-1}cm^{2}
\eeq
The $\tau$ lepton may decay or undergo charged current interaction to produce
 a $\nu_{\tau}$. The $\nu_{\tau}$ produced from $\tau$ decay carries $40\%$ of the $\tau$ lepton energy.
The expression for the probability of neutral current interaction including $\tau$ decay and $\tau$ charged current interaction is
\bea
P_{\nu_{\tau}\rightarrow \nu_{\tau}}(E_{\nu_{\tau}}(0),E_{\nu_{\tau}}(x),x)
dE_{\nu_{\tau}}(x)&=&P_{\nu_{\tau}\rightarrow \nu_{\tau}}^{0}(E_{\nu_{\tau}}(0),E_{\nu_{\tau}}(x),x) dE_{\nu_{\tau}}(x)\nonumber\\
&+& dE_{\nu_{\tau}}(x)\int_{0}^{x}\int_{0}^{E_{\tau}(x_1)}
P_{\nu_{\tau}\rightarrow \nu_{\tau}}^{0}(E_{\nu_{\tau}}(x_1),E_{\nu_{\tau}}(x),x-x_1)\nonumber\\
&\times&(\frac{d\Gamma_{\tau}(E_{\nu_{\tau}}(x_1),E_{\tau}(x_1))}{dE_{\nu_{\tau}}(x_1)}\nonumber\\&+&\frac{d\sigma_{cc}^{\tau N \rightarrow \nu_{\tau} X}(E_{\nu_{\tau}}(x_1),E_{\tau}(x_1))}{dE_{\nu_{\tau}}(x_1)}\rho(x_1)N_A)
dE_{\nu_{\tau}}(x_1) dx_1
\nonumber\\
&\times&\int_0^{x_1}\int_{0}^{E_{\nu_{\tau}}(x_2)}P_{\nu_{\tau}\rightarrow {\tau}}(E_{\nu_{\tau}}(0),E_{\tau}(x_2),x_2)dE_{\tau}(x_2)dx_2\nonumber\\ 
\eea

The energy of the $\tau$ lepton at $x_1$ and $x_2$ are related by the energy loss equation of $\tau$ leptons.
\beq
\frac{dE_{\tau}(x)}{dx}=-\beta(E_{\tau}(x))E_{\tau}(x)
\eeq
The expression for the probability of charged current interaction is
\bea
P_{\nu_{\tau}\rightarrow {\tau}}(E_{\nu_{\tau}}(0),E_{\tau}(x),x)dE_{\tau}(x)dx&=&\rho(x) N_A dE_{\tau}(x)dx \nonumber\\
&\times& \int_{E_{{\nu}min}}^{E_{{\nu}max}}P_{\nu_{\tau}\rightarrow \nu_{\tau}}(E_{\nu_{\tau}}(0),E_{\nu_{\tau}}(x),x)\nonumber\\
&\times&\frac{d\sigma_{CC}(E_{\nu_{\tau}}(x),E_{\tau}(x))}{dE_{\tau}(x)}dE_{\nu_{\tau}}(x). 
\eea
The $\tau$ leptons produced by charged current interaction lose energy before
 emerging from the surface of the Earth. The range of $\tau$ is approximately $R_{\tau}=1/{\beta(E_{\tau}(x))} \ln(E_{\tau}(x)/E_{th})$ from eqn.(24). $E_{th}$ is the threshold energy of the detector. 
The range of tau leptons can be written as a function of $E_{\nu_{\tau}}(x)$
by replacing $E_{\tau}(x)$ with $0.75E_{\nu_{\tau}}(x)$.
The final expression for the probability of emission of $\tau$ leptons from the surface of the Earth having energies above the threshold energy of the detector is obtained after integrating eqn.(25) over $E_{\tau}(x)$ and $x$.
\beq
Prob({\nu_{\tau}}\rightarrow {\tau})=\int_{0}^{R_{\tau}}P_{{\nu_{\tau}}\rightarrow {\tau}}(E_{\nu_{\tau}}(0),0.75E_{\nu_{\tau}}(x),x)dx 
\eeq
We have solved eqn.(23) and (25) iteratively to find the value of the probability on the left hand side of eqn.(25). We use the charged current and neutral current interaction cross-sections from \cite{raj} and the Earth's density profile from \cite{earth}. Our result converges after two iterations. Finally we have use the probability calculated in eqn.(26)
 $Prob({\nu_{\tau}}\rightarrow {\tau})$ for our calculation of number of $\tau$ events in the detector. 


\newpage
\begin{thebibliography} {99}
\bibitem{sup}http://www-sk.icrr.u-tokyo.ac.jp/doc/sk/.
\bibitem{kam}http://www.awa.tohoku.ac.jp/html/KamLAND.
\bibitem{sno}http://www.sno.phy.queensu.ca.
\bibitem{beacom1}J. F. Beacom, N. Bell, D. Hooper, S. Pakvasa and T. J. Weiler,
Phys. Rev. Lett. {\bf 90}, 181301 (2003).
\bibitem{beacom2}J. F. Beacom, N. F. Bell, D. Hooper, J. G. Learned, S. Pakvasa
and T. J. Weiler, Phys. Rev. Lett. {\bf 92}, 011101 (2004).
\bibitem{ice}http://icecube.wisc.edu/
\bibitem{anita}http://www.ps.uci.edu/~anita/.
\bibitem{owl}F.Krizmanic et al. [OWL/AirWatch Collaboration], in Proceedings
of the ${26}^{th}$ International Cosmic Ray Conference (ICRC 99), Salt Lake 
City 1999, Vol. 2, 388-391.
\bibitem{waxman}E. Waxman, J. Bahcall, Phys.Rev.Lett. {\bf78}, 2292 (1997).
\bibitem{nayan1}N. Gupta, Phys. Rev. D {\bf 65}, 113005 (2002).
\bibitem{guetta}D. Guetta, D. Hooper, J. Alvarez-Mu\~niz, F. Halzen and E. Reuveni, Astropart. Phys. {\bf 20}, 429 (2004). 
\bibitem{owl1}For Observation of Neutrino Universe with OWL-AIRWATCH, 
OWL/AIRWATCH Science Working Group, D. Cline et al (1999);
 D. B. Cline, F. W. Stecker, astro-ph/0003459.  
\bibitem{nayan}N. Gupta, Phys.Rev. D {\bf 68}, 063006 (2003).
\bibitem{kolb}E. W. Kolb and M. S. Turner, The Early Universe, {\it ADDISON-WESLEY PUBLISHING COMPANY 1990}.
\bibitem{wmap} C. L. Bennett et al., ApJ. S. {\bf 148}, 1 (2003).
\bibitem{hal}F. Halzen and D. Saltzberg, Phys. Rev. Lett. {\bf 81}, 4305 (1998).
\bibitem{beacomr}J. F. Beacom, P. Crotty and E. W. Kolb, Phys. Rev. D {\bf 66},
021302(R) (2002). 
\bibitem{naumov}V. A. Naumov and L. Perrone, Astropart. Phys. {\bf 10}, 239 (1999).
\bibitem{sharada}S. I. Dutta, M. H. Reno, I. Sarcevic, Phys. Rev. D {\bf 62},
 123001 (2000).
\bibitem{bottai}S. Bottai and S. Giurgola, Astropart. Phys. {\bf 18}, 539 (2003).
\bibitem{tseng}J. -J. Tseng, T. -W. Yeh, H. Athar, M. A. Huang, F. -F. Lee and G. -L. Lin, Phys. Rev. D {\bf 68}, 063003 (2003).
\bibitem{bugaev}E. Bugaev, T. Montaruli, Y. Shlepin and I. Sokalski, 
astro-ph/0312295, Astropart. Phys. in press.
\bibitem{yoshida}S. Yoshida, R. Ishibashi and H. Miyamoto, astro-ph/0312078, Phys. Rev. D in press.
\bibitem{owl_n}F. W. Stecker et al., astro-ph/0408162.
\bibitem{plb}We thank P. L. Biermann for this suggestion.
\bibitem{wax1}E. Waxman, astro-ph/0103186, Based on lectures given at the ICTP Summer School (ICTP, Italy, June 2000), and at the VI Gleb Wataghin School (UNICAMP, Brazil, July 2000.
\bibitem{kim}Neutrinos in Physics and Astrophysics, C. W. Kim and A. Pevsner,
Harwood Academic, 1993.
\bibitem{amit}HSM would like to thank A. Dutta for this observation.
\bibitem{raj}R. Gandhi, C. Quigg, M. H. Reno and I. Sarcevic, Phys. Rev. D {\bf 58}, 093009 (1998).
\bibitem{earth}http://pubs.usgs.gov/gip/interior/.
\end{thebibliography}
\end{document}